# Synthesis and properties of $A_xV_2Al_{20}$ ($A$ = Th, U, Np, Pu) ternary actinide aluminides

M. J. Winiarski[a],[*] J.-C. Griveau[b], E. Colineau[b], K. Wochowski[c], P. Wiśniewski[c],

D. Kaczorowski[c], R. Caciuffo[b], T. Klimczuk[a],[†],

[a] *Faculty of Applied Physics and Mathematics, Gdansk University of Technology, Narutowicza 11/12, 80-233 Gdansk, Poland*
[b] *European Commission, Joint Research Centre (JRC), Directorate for Nuclear Safety and Security, Postfach 2340, 76125 Karlsruhe, Germany*
[c] *Institute of Low Temperature and Structure Research, Polish Academy of Sciences, PO Box 1410, 50-950 Wrocław, Poland*

* *corresponding author: mwiniarski@mif.pg.gda.pl*

† *-corresponding author: tomasz.klimczuk@pg.gda.pl*

**Abstract**

Polycrystalline samples of $A_xV_2Al_{20}$ ($A$ = Ce, Th, U, Np, Pu; $0.7 \leq x \leq 1.0$) actinide intermetallics were synthesized using the arc-melting method. Crystal structure studies were performed by means of powder x-ray diffraction and the Rietveld refinement method. All studied compounds crystallize in the $CeCr_2Al_{20}$-type structure (space group Fd-3m, no. 227) with the actinide or Ce atom located in the oversized cage formed by Al atoms. Comparison of the crystallographic results with the reported data for $LnV_2Al_{20}$ (Ln = lanthanoids) counterparts reveals distinctly different behavior of the lanthanide- and actinide-bearing compounds. This difference is suggested to be caused by fairly localized character of the 4*f* electrons, whereas itinerant character of the 5*f* electrons is likely seen for U- and Np-containing phases. Magnetic susceptibility and specific heat measurements did not reveal any magnetic ordering in $U_{0.8}V_2Al_{20}$, $Np_{0.8}V_2Al_{20}$ and $Pu_{0.8}V_2Al_{20}$ down to 2.5 K. A small anomaly in low-temperature specific heat of $Ce_{0.8}V_2Al_{20}$, $U_{0.8}V_2Al_{20}$, and $Np_{0.8}V_2Al_{20}$ is observed, likely arising from a low-energy Einstein mode.

## 1. Introduction

The $CeCr_2Al_{20}$-type aluminide family have attracted much attention due to the observation of various physical phenomena, including antiferromagnetic ordering [1–4], itinerant ferromagnetism [5,6] and heavy fermion behavior [2,5,7]. Several $CeCr_2Al_{20}$-type superconductors were also reported [8–12], some of which show coexistence of the superconducting state with quadrupolar magnetic ordering [10,11]. Recently, it was shown that the so-called "rattling" effect can enhance the superconducting transition temperature in $RV_2Al_{20}$, where $R$ is a non-magnetic rare earth (Sc, Lu, Y) [12].

The group of $CeCr_2Al_{20}$-type intermetallics was discovered in the late 1960's. [13,14]. In the unit cell of $CeCr_2Al_{20}$ (face-centered cubic system, sg. *Fd-3m*, no. 227), Ce (8*a*) atoms are positioned in cages formed by Al atoms, while Cr (16*d*) atoms form a pyrochlore lattice (see Fig. 1(a)).

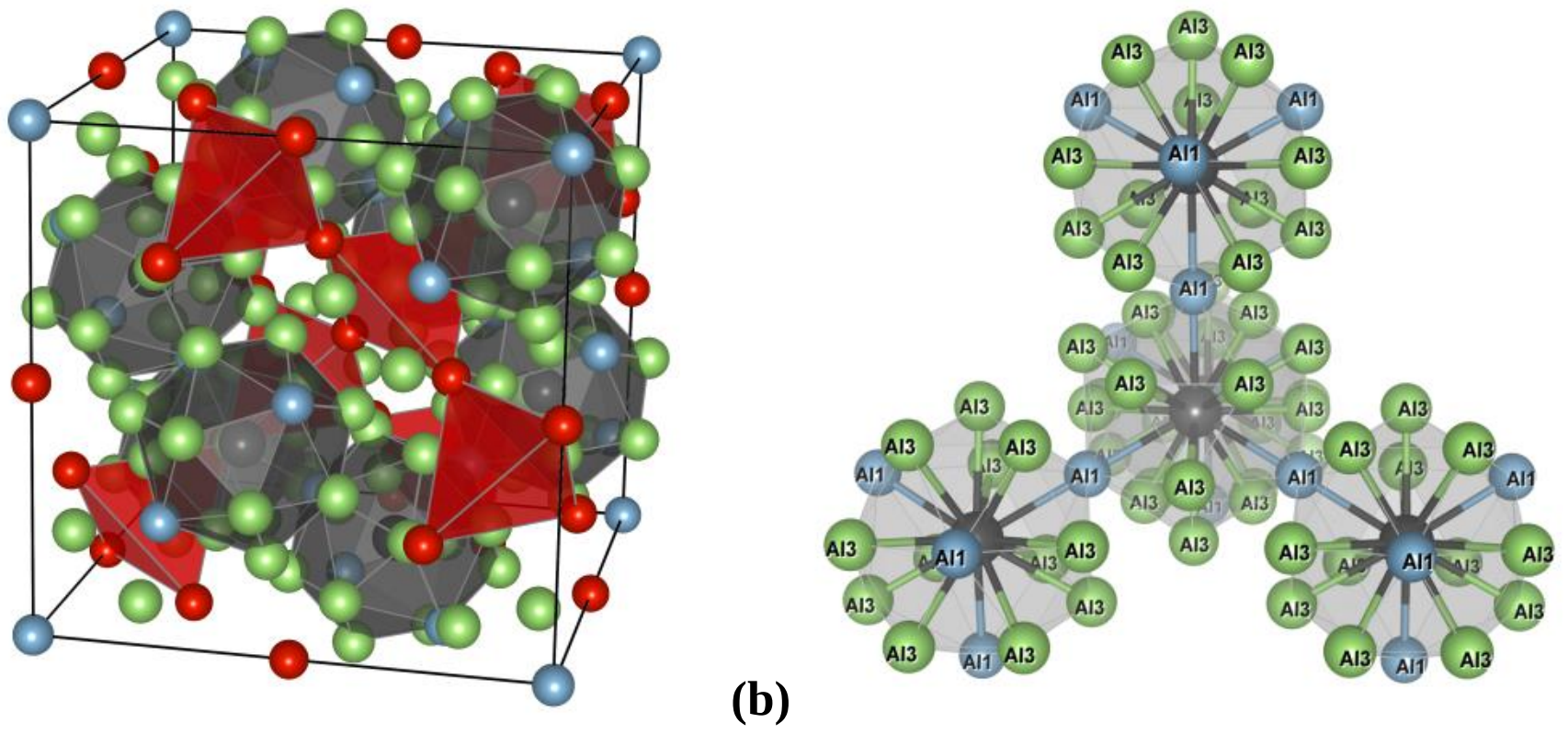

(a) (b)

Fig. 1. (a) The unit cell of $CeCr_2Al_{20}$-type $Pu_{0.8}V_2Al_{20}$ compound. Pu atoms (grey) occupy 8a (⅛,⅛,⅛) position and are surrounded by 16 Al atoms (green, blue), forming an oversized cage. Cr atoms (red) are located at 16*d* (½,½,½) site, and form a pyrochlore lattice. Al atoms occupy 16*c*, 48*f*, and 96*g* positions. (b) Local structure of Pu atoms (grey) in $Pu_{0.8}V_2Al_{20}$ compound. The Pu atom is surrounded by 16 Al atoms: 4 Al1(16*c*), blue and 12 Al3(96*g*), green. The neighboring $PuAl_{16}$ polyhedra share one Al1(16*c*) atom. Note that Al1-Pu and Al3-Pu distances may differ significantly (see Table 1). The image was rendered using VESTA [15].

The family of $CeCr_2Al_{20}$-type compounds include more than 80 intermetallics known to date, based on Al, Zn, and Cd, with different 3*d*, 4*d*, and 5*d* metals in Cr position and a variety of electropositive elements replacing Ce (Ca, rare earth metals, Th, U, Zr, Hf, Nb, Al, and Ga). The reports on actinide-containing $CeCr_2Al_{20}$-type compounds are inevitably scarce and limited to Th and U, due to the radioactivity and low availability of heavier actinide elements. Halevy *et al.* [16] reported the synthesis and characterization of $UV_2Al_{20}$. X-ray diffraction (XRD) and scanning electron microscope investigations have shown, however, that the sample contained some amount of $UAl_4$ and Al-U solid solution. This suggested that the real composition could be slightly sub-stoichiometric. Recently Uziel *et al.* [17] studied $ThT_2Al_{20}$ ($T$ = Ti-Fe) materials, revealing that the $CeCr_2Al_{20}$-type phase is obtained only with $T$ = Ti, V, Cr. The only known Mn-bearing $CeCr_2Al_{20}$-type aluminide, $UMn_2Al_{20}$, was established to exhibit an itinerant ferromagnetic behavior [5,6].

In this study we report the results on synthesis and crystal structure studies of $AV_2Al_{20}$ ($A$ = Ce, Th, U, Np, Pu) intermetallics along with the physical properties of the selected compounds, including the two first $CeCr_2Al_{20}$-type compounds containing transuranium elements, namely $Np_{0.8}V_2Al_{20}$ and $Pu_{0.8}V_2Al_{20}$.

## 2. Materials and Methods

Polycrystalline samples of the materials studied were prepared by arc-melting using vanadium (purity 99.8%) and aluminum (purity 99.999%) as reagents. Three $Th_xV_2Al_{20}$ (x = 0.8-1.0) samples were prepared with thorium (purity 99.6%). The synthesis was performed in three steps: (1) preparation of $VAl_3$ binary phase, (2) melting of the obtained alloy with additional aluminum quantity appropriate to obtain $VAl_{10}$, and (3) melting the $VAl_{10}$ sample with thorium. The resulting button was remelted several times to ensure good homogeneity.

$U_xV_2Al_{20}$ (x = 0.7-1.0) and $Ce_xV_2Al_{20}$ (x = 0.5-1.0) were synthesized using uranium (purity 99.8%) or cerium (purity 99.9%) metals, which in the first step were melted with V and approximately one third of the total Al content, and then remelted with appropriate amount of Al to obtain $A_xV_2Al_{20}$ stoichiometry.

In the case of $Np_xV_2Al_{20}$ (x = 0.7-0.8) and $Pu_{0.8}V_2Al_{20}$, binary aluminides $NpAl_2$ [18] and $PuAl_2$ [19] were taken as precursors. The melting process involved multiple steps: (1) preparation of $VAl_3$ binary compound, (2) melting $VAl_3$ with $NpAl_2$ and $PuAl_2$, (3) adding Al (in two steps), and (4) gently remelting three times to promote homogeneity. Arc melting was performed in high purity argon atmosphere in an arc-furnace placed in radiation-protective glove box filled with neutral gas (nitrogen).

As cast samples were wrapped in tantalum foil, sealed in evacuated quartz tubes and annealed for 3 (Np, Pu) or 4 weeks (Ce, Th, U) at 645°C. This annealing temperature was chosen due to the fact that the melting temperature of pure Al metal is $661^oC$.

Crystallographic structure of Np- and Pu-containing samples were determined by powder x-ray diffraction (PXRD) at room temperature on a Bruker D8 Focus diffractometer equipped with a Cu $K_{\alpha 1}$ source and a germanium (111) monochromator running in a glove box for measurements on transuranium compounds. To avoid contamination of the glove box, the powdered specimens were embedded in epoxy glue. The Ce-, Th- and U-containing samples were examined using X'Pert PRO powder x-ray diffractometer with Cu $K_\alpha$ radiation. The structural parameters were refined by means of the Rietveld method [20] using the FullProf package [21].

Magnetic susceptibility measurements on $U_{0.8}V_2Al_{20}$, $Np_{0.8}V_2Al_{20}$ and $Pu_{0.8}V_2Al_{20}$ were conducted using a Quantum Design MPMS superconducting quantum interference device (SQUID) magnetometer. Samples were held in Plexiglas tubes. Raw longitudinal magnetization data have been corrected for diamagnetic contributions and small additional signal due to sample holder. The magnetization of $U_{0.8}V_2Al_{20}$ and $Np_{0.8}V_2Al_{20}$, was measured from 2 to 300 K. In the case of $Pu_{0.8}V_2Al_{20}$, due to radiation heating, the lowest achieved temperature was 3.5 K.

Specific heat measurements were done employing a Quantum Design PPMS-9 platform using the standard 2τ-relaxation method. The experiments were performed from 300 K down to 1.9 K ($Ce_{0.8}V_2Al_{20}$), 2.0 K ($U_{0.8}V_2Al_{20}$), 2.1 K ($Np_{0.8}V_2Al_{20}$), 2.5 K ($Pu_{0.8}V_2Al_{20}$), and 0.4 K ($Th_{0.8}V_2Al_{20}$). $Np_{0.8}V_2Al_{20}$ and $Pu_{0.8}V_2Al_{20}$ samples were encapsulated in the heat conducting STYCAST 2850 FT resin to prevent the release of radioactive material [22,23]. The heat capacity data were corrected for the contribution of the encapsulation material by using an empirical relation determined previously [24].

## 3. Results and discussion

Powder x-ray diffraction patterns indicated fairly high quality of the annealed samples with concentration of the $A$ atom x = 0.8. Stoichiometric $AV_2Al_{20}$ ($A$ = Ce, Th, U) compounds have not been obtained, which suggests that some substoichiometry (x < 1) is required in order to get chemically pure samples. This finding is in line with the previous report on $UV_2Al_{20}$ [16]. We found that nominally stoichiometric $CeV_2Al_{20}$ and $ThV_2Al_{20}$ contained small amount of an unidentified impurity, possibly binary Ce-Al and Th-Al compounds, respectively. Some impurity peaks were observed in the XRD pattern of $Th_{0.8}V_2Al_{20}$ that can be assigned to the monoclinic $V_7Al_{45}$ binary phase [25]. An effort to synthesize pure $U_xV_2Al_{20}$ with x

< 0.7 was unsuccessful, with a secondary phase being most likely the $VAl_{10+x}$ compound. The two Np-bearing samples contained a trace of an unrecognized impurity phase, with $Np_{0.7}V_2Al_{20}$ exhibiting a significantly worse quality.

The crystallographic parameters derived from the Rietveld refinements of the structural model to match the powder x-ray diffraction data for $A_{0.8}V_2Al_{20}$ ($A$ = Ce, Th, U, Np, Pu) are gathered in Table 1. Plots of the fits are shown in the Supplementary Material (Figs. S1-S5). The fit reliability factors ($R_p$, $R_{wp}$ and $R_{exp}$) estimated for Np-bearing sample are much higher than for the other materials, suggesting worse quality of the structural model, that should be therefore considered as only a rough approximation of the real structure. A comparison of the cell parameter dependence on the covalent radius of the cage-filling atom (Figure 2) reveals an essential difference between the lanthanide- and actinide-bearing $A_xV_2Al_{20}$ compounds. A positive relation is observed for (La, Pr-Lu)$V_2Al_{20}$ system as opposed to a negative trend observed for (U, Np, Pu)$V_2Al_{20}$. This behavior can be rationalized by the tendency of 5*f* electrons towards delocalization, as opposed to the mostly localized character of lanthanide 4*f* states. In the latter, the atomic size is the main parameter governing the unit cell size, since most of the lanthanides share the +3 valency. In case of the former, the delocalized 5*f* electrons contribute to the bonding, resulting in more compact structures. As the itinerancy of the 5*f* electrons decreases with increasing atomic number the two trends converge for $Pu_{0.8}V_2Al_{20}$, in agreement with a common belief that Pu locates at a border of a localized-itinerant transition [26].

Both $Ce_{0.8}V_2Al_{20}$ and $Th_{0.8}V_2Al_{20}$ compounds deviate from the trend forming a third group (see Fig. 2(a)). This phenomenon is likely caused by the +4 valency of Ce and Th, as one more electron is available for bonding, leading to smaller unit cell, compared to trivalent lanthanides. The tetravalent (nonmagnetic) nature of Ce in $CeV_2Al_{20}$ was previously observed in magnetic susceptibility measurements [27,28]. Figure 2(b) presents the cubic lattice parameter versus occupancy of the cage atom for $Ce_xV_2Al_{20}$, $Th_xV_2Al_{20}$ and $U_xV_2Al_{20}$. A positive linear behavior is observed for both $Ce_xV_2Al_{20}$ and $Th_xV_2Al_{20}$ with the last points (x = 1.0) deviating from the trend. In contrary, for $U_xV_2Al_{20}$ the lattice parameter changes only slightly, decreasing by just ca. 0.001 Å from x = 0.7 to 0.9. Stoichiometric (x=1.0) samples have not been obtained.

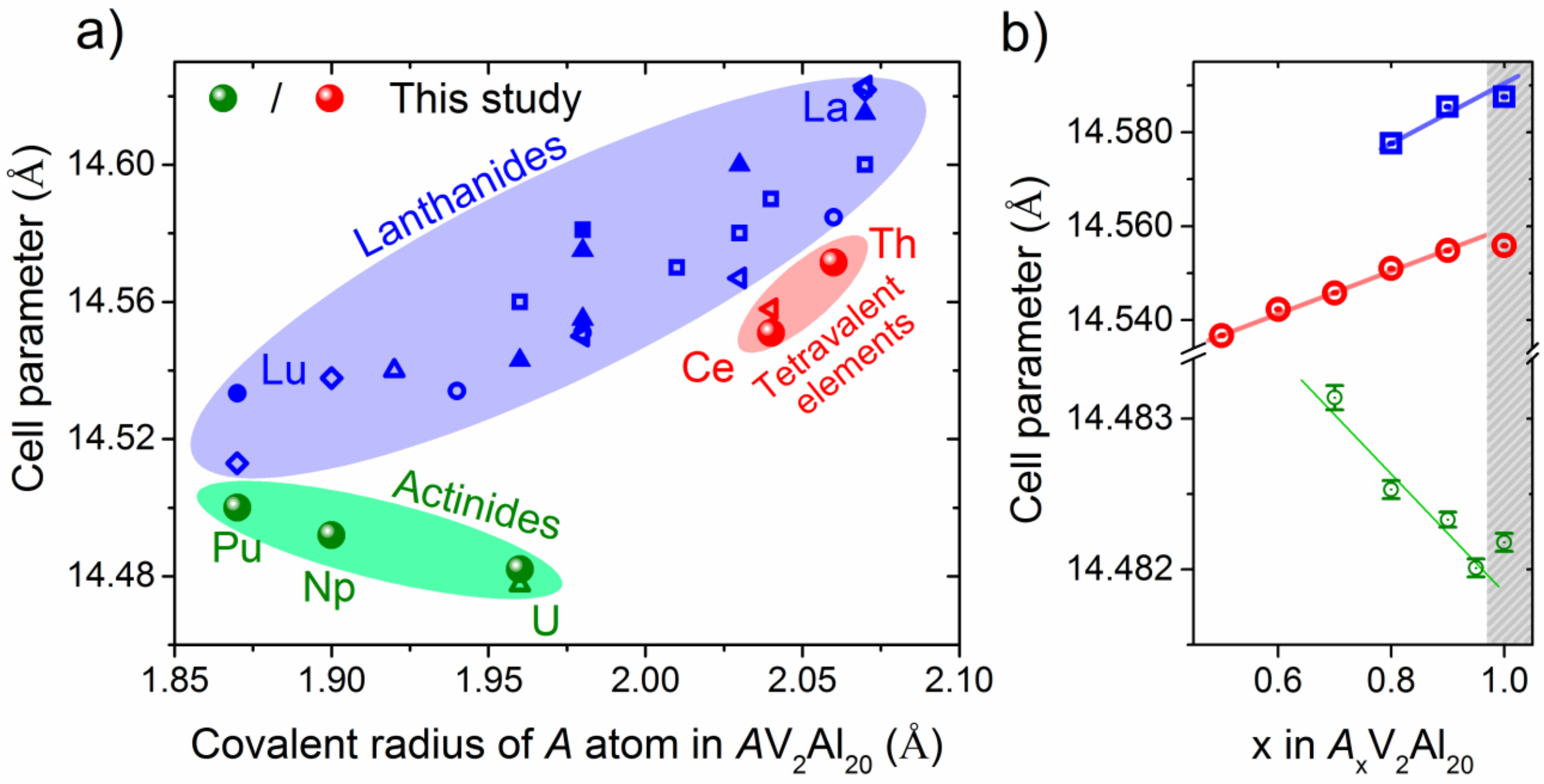

Fig. 2. (a) Relation between a lattice constant and radii of the cage-filling $A$ atom in $AV_2Al_{20}$ intermetallic system. Three different groups are noticeable. The data are taken from refs [12–14,16]. Samples marked by filled balls are discussed in the manuscript and have $A_{0.8}V_2Al_{20}$ stoichiometry. The covalent radii are used as a measure of atom sizes, not as an indication of bonding character. Note however, that the metallic radii of lanthanides follow the same trend (lanthanide contraction), and in case of the actinides, the metallic radii also decrease with increasing atomic number given the same number of localized electrons [29,30]. (b) The dependence of the cell parameter versus the concentration of the $A$ atom in $Th_xV_2Al_{20}$, $Ce_xV_2Al_{20}$, and $U_xV_2Al_{20}$. The negative dependence observed for the U series is much weaker than the positive trend seen for Th and Ce series. Solid lines are guides for eyes and a shadow region marks a multiphase region.

Table 1. Crystallographic structure parameters of $A_{0.8}V_2Al_{20}$ ($A$ = Ce, Th, U, Np, Pu) obtained from Rietveld fits to powder diffraction data. Volumes of the *A*-Al cages are calculated using VESTA software [15]. Covalent radii are given after Cordero *et al.* [31]. The *R*-factors given in the table are conventional Rietveld refinement reliability factors (corrected for background), calculated only for points with Bragg contributions. Reliability factors calculated without correction for background are given in Table S1 of Supplementary Material. For discussion of effect of background in fit reliability assessment see refs. [32,33].

| | **$Ce_{0.8}V_2Al_{20}$** | **$Th_{0.8}V_2Al_{20}$** | **$U_{0.8}V_2Al_{20}$** | **$Np_{0.8}V_2Al_{20}$** | **$Pu_{0.8}V_2Al_{200}$** |
|---|---|---|---|---|---|
| Space group | *Fd*-3*m* (# 227) | | | | |
| Pearson symbol | *cF*184 | | | | |
| *Z* (number of formula units per unit cell) | 8 | | | | |
| Cell parameter (Å) | 14.5510(1) | 14.5782(1) | 14.4821(2) | 14.4920(1) | 14.5000(1) |
| Cell volume (Å$^3$) | 3080.907 | 3098.210 | 3037.349 | 3043.594 | 3048.625 |
| Molar weight (g·mol$^{-1}$) | 753.6058 | 827.1435 | 831.9361 | 831.1130 | 836.7130 |
| Density (g·cm$^{-3}$) | 3.295 | 3.547 | 3.639 | 3.628 | 3.646 |
| *A* (8*a*) x = y = z = | 1/8 | | | | |
| V (16*d*) x = y = z = | ½ | | | | |
| Al1 (16*c*) x = y = z = | 0 | | | | |
| Al2 (48*f*) x = | 0.4872(1) | 0.4885(1) | 0.4873(1) | 0.4909(2) | 0.4864(1) |
| y = z = | 1/8 | | | | |
| Al3 (96*g*) x = y = | 0.0592(1) | 0.0589(1) | 0.0592(1) | 0.0589(1) | 0.0594(1) |
| z = | 0.3250(1) | 0.3258(1) | 0.3242(1) | 0.3239(2) | 0.3243(1) |
| *A*-*A* distance (Å) | 6.30 | 6.31 | 6.27 | 6.28 | 6.28 |
| Hill limit (Å) | 3.4 | - | 3.4-3.6 | 3.25 | 3.4 |
| *A*-Al distances: *A*-Al1 (16*c*): | 3.209 | 3.229 | 3.184 | 3.185 | 3.187 |
| *A*-Al3 (96*g*): | 3.150 | 3.156 | 3.135 | 3.138 | 3.139 |
| Average *A*-Al(16*c*/96*g*) distance, $R_{cage}$ (Å) | 3.19 | 3.21 | 3.17 | 3.17 | 3.18 |
| *A*-Al cage volume, $V_{cage}$ (Å$^3$) | 93.84 | 95.27 | 91.87 | 92.00 | 92.18 |
| Covalent radius of *A* atom, $R_{atom}$ (Å) | 2.04 | 2.06 | 1.96 | 1.90 | 1.87 |
| $R_{atom}/R_{cage}$ | 63.9% | 64.2% | 61.8% | 59.9% | 58.8% |
| $V_{atom}/V_{cage}$ | 37.9% | 38.4% | 34.3% | 31.2% | 29.7% |
| Refinement reliability factors: | | | | | |
| $R_p$ (%) | 11.9 | 12.5 | 9.14 | 23.4 | 15.4 |
| $R_{wp}$ (%) | 9.26 | 8.40 | 9.42 | 16.8 | 9.83 |
| $R_{exp}$ (%) | 5.02 | 6.42 | 4.52 | 11.4 | 7.47 |
| $\chi^2$ (%) | 3.40 | 1.71 | 4.34 | 2.16 | 1.73 |

The $A_{0.8}V_2Al_{20}$ ($A$ = Ce, Th, U, Np, Pu) samples were selected for further characterization. Both the transuranium element-bearing samples show a linear weakly temperature dependent character of the magnetic susceptibility χ(T), as it is displayed in Figure 3 (a). Previous studies on $UT_2Al_{20}$ ($T$ = Ti, Cr, Nb, Mo, and W) have also revealed a small and nearly temperature-independent magnetic susceptibility that signaled itinerant character of U-5*f* electrons [6,34–36]. The magnetic susceptibility of $Np_{0.8}V_2Al_{20}$ is smaller than that of $Pu_{0.8}V_2Al_{20}$ resulting from lower density of states at the Fermi energy DOS($E_F$), which is proportional to the Pauli contribution to temperature-independent susceptibility. $U_{0.8}V_2Al_{20}$, however, does not fit this scheme, as the susceptibility is ca. twice larger than in Pu compound, while lower DOS($E_F$) is expected. This could be explained by the presence of a paramagnetic impurity. Alternatively, the enhanced susceptibility may arise from the difference in the strength of electron correlations. The presence of exchange interactions lead to renormalization of susceptibility as compared to the free-electron case.

For each compound, no magnetic ordering is observed down to the lowest temperatures studied. Figure 3 (b) shows the magnetization of $U_{0.8}V_2Al_{20}$ measured at 5 K as a function of the magnetic field. A linear dependence up to 7 T, without hysteresis, corroborates a Pauli paramagnetic character of the compound. The weakly temperature dependent susceptibility of $U_{0.8}V_2Al_{20}$ could not be modelled with modified Curie-Weiss law and may arise from a minor contamination with paramagnetic impurities.

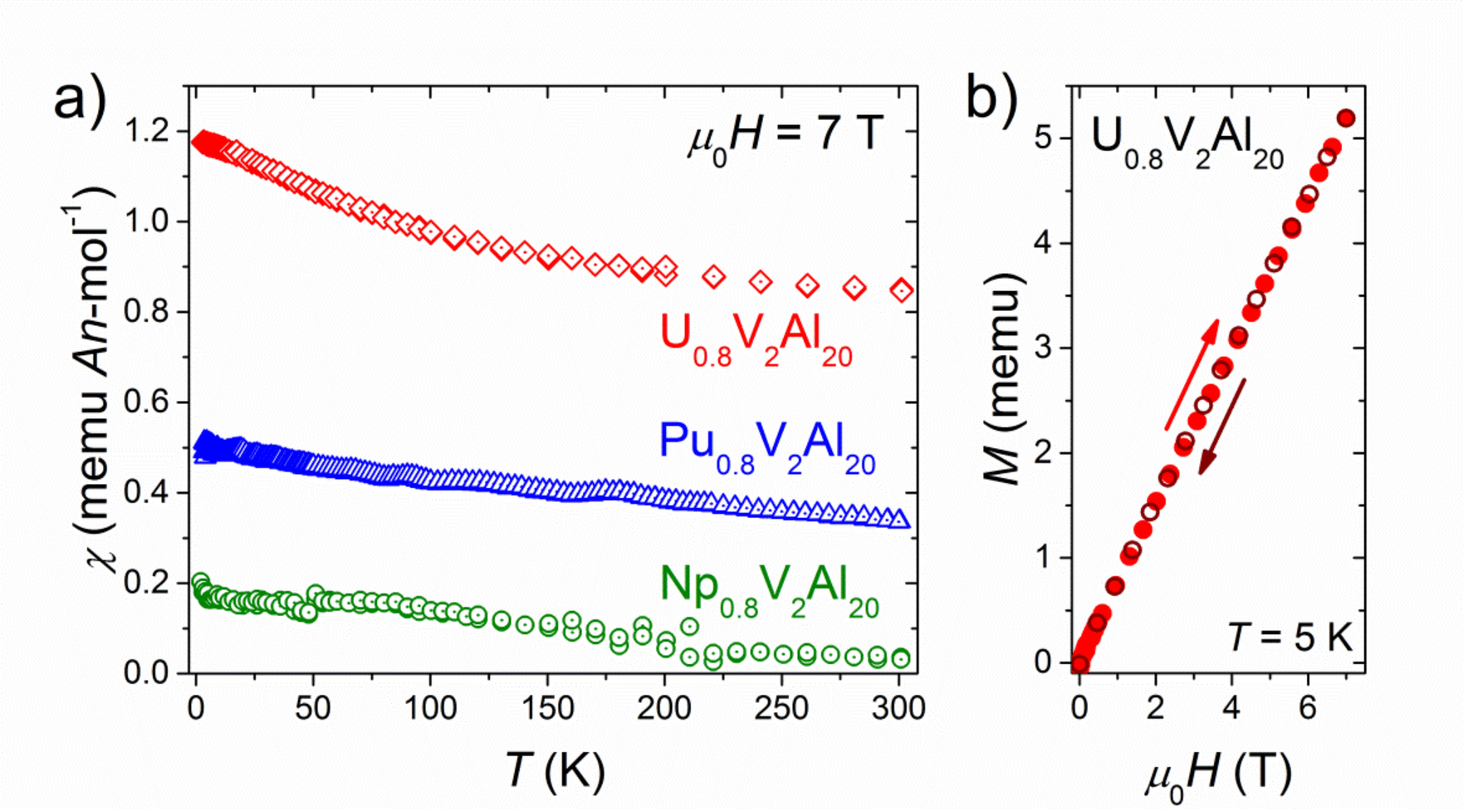

Fig. 3. (a) Temperature dependence of the magnetic susceptibilities χ(T) of $U_{0.8}V_2Al_{20}$, $Np_{0.8}V_2Al_{20}$, and $Pu_{0.8}V_2Al_{20}$ per mole of actinide atom at 7 T magnetic field. No sign of magnetic ordering is observed (b) Plot of magnetization vs. applied field $M(H)$ for $U_{0.8}V_2Al_{20}$ displaying a linear behaviour up to 7 T field.

The lack of local magnetic moment in $A_{0.8}V_2Al_{20}$ ($A$ = U, Np, Pu) suggests delocalization and thus strong hybridization of the 5*f* states. The 5*f* delocalization is possible via two hybridization mechanisms: direct *f*-*f* or *f*-ligand [37]. The former is typically observed when the separation between the actinide atoms is below the Hill limit. In our case, the inter-actinide distance is almost twice the Hill limit, and using simple reasoning, one should expect to observe a localized magnetic moment on *A* atoms. However, the assumption behind Hill plots is that the *f*-*f* orbital overlap is the only parameter governing the 5*f* hybridization, and thus it does not account for the *f*-ligand hybridization caused by overlap with orbitals of neighboring atoms.

The *f*-ligand hybridization is expected to be strong in cases where the coordination of *f* atoms is large [37]. In the $CeCr_2Al_{20}$ aluminides the 8*a* atom is surrounded by as many as 16 Al atoms with an average *A*-Al distance of around 3.2 Å in case of the actinide-bearing compounds (see Table 1), making the structure favorable for delocalized *f* electron behavior. In fact, the Ce 4*f* states in $CeT_2Al_{20}$ ($T$ = Ti, V, Cr) are found to show a highly hybridized character compared to isostructural $CeIr_2Zn_{20}$, in which valence instability is observed [38] and finally $CeT_2Cd_{20}$ ($T$ = Ni, Pd) in which Ce 4*f* hybridization is weak and electrons are fairly localized [39]. Similarly, in $UT_2Al_{20}$ ($T$ = Ti, Cr, Nb, Mo, W) the 5*f* electrons are strongly delocalized [6,34–36] while in $UPd_2Cd_{20}$ the large local magnetic moment suggests only a weak hybridization [40], with Zn-based compounds [41] falling in between the two regimes.

The two observations: that delocalization is stronger for Al-based than for Cd- and Zn-based $CeCr_2Al_{20}$-type compounds, and that this trend holds both for Ce and U, are in agreement with general rules for *f*-ligand hybridization summarized by Koelling *et al.* [37] who pointed out that firstly, the character of the ligand atom is primary in determining the *f* states behaviour, leading to similarities between Ce and U-bearing intermetallic compounds, and secondly, that the hybridization affects strongly only early lanthanides and actinides, due to larger spatial extent of the *f* wavefunction than in heavier elements.

Considering the $CeCr_2Al_{20}$-type crystallographic structure and taking into account that the large spatial separation between the 5*f* atoms (over 6 Å) prevents the direct hybridization due to lack of 5*f*-5*f* overlap, one may conclude that the hybridization between the 5*f* states and the orbitals of neighboring Al atoms is significant and leads to strongly itinerant character.

It is worth noting that the *f*-ligand hybridization is said to be responsible for lack of 5*f* moment in $AuCu_3$-type actinide compounds: $UM_3$ ($M$ = Ge, Rh, Ir) (see ref. [37] and references therein) and $NpM_2$ ($M$ = Ge, Rh) [42–44]. In these cases, the evidence for the 5*f*-ligand hybridization scenario comes from both experimental and theoretical investigations [37,42,43,45].

The specific heat of $A_{0.8}V_2Al_{20}$ ($A$ = Th, U, Np, Pu) is shown in Figure 4 in the form of a $C_p/T$ vs. $T^2$ plot. The low-temperature data was fitted using a relation shown in eq. 1:

$$\frac{C_p}{T} = \gamma + \beta T^2 \qquad (1)$$

where γ is the Sommerfeld electronic heat capacity coefficient and the second term accounts for the phonon contribution. The obtained values of γ significantly vary between the four compounds, ranging from 25 to 88 mJ $mol^{-1}$ $K^{-2}$ for Th- and Np-bearing sample, respectively. The lowest γ is similar to the estimated γ = 26.5 mJ $mol^{-1}K^{-2}$ for $ScV_2Al_2$ [12]. The moderately high value of γ for $Np_{0.8}V_2Al_{20}$ and $U_{0.8}V_2Al_{20}$ suggests moderate electron-electron correlations. The γ value for $U_{0.8}V_2Al_{20}$ is close to 60 mJ $mol^{-1}$ $K^{-2}$ observed in $UNb_2Al_{20}$ [36] and lower than 80 mJ $mol^{-1}$ $K^{-2}$ reported for $UCr_2Al_{20}$ compound [34] in which contribution of the Cr 3*d* electrons at the Fermi level is responsible for Sommerfeld coefficient enhancement. The contribution of Cr states at $E_F$ is also responsible for larger value of γ observed for $ThCr_2Al_{20}$ (62 mJ $mol^{-1}$ $K^{-2}$) [34] compared to $Th_{0.8}V_2Al_{20}$.

No sign of nuclear Schottky anomaly, commonly found for the Np-based intermetallic compounds well above 2 K [46–49], was observed in $C_p(T)$, suggesting that the magnetic hyperfine fields acting on Np nuclei are weak. This is in agreement with the delocalized character of Np 5*f* electrons. The Schottky anomaly may be observed in yet lower temperatures due to the quadrupole splitting of Np ground state by crystal electric field.

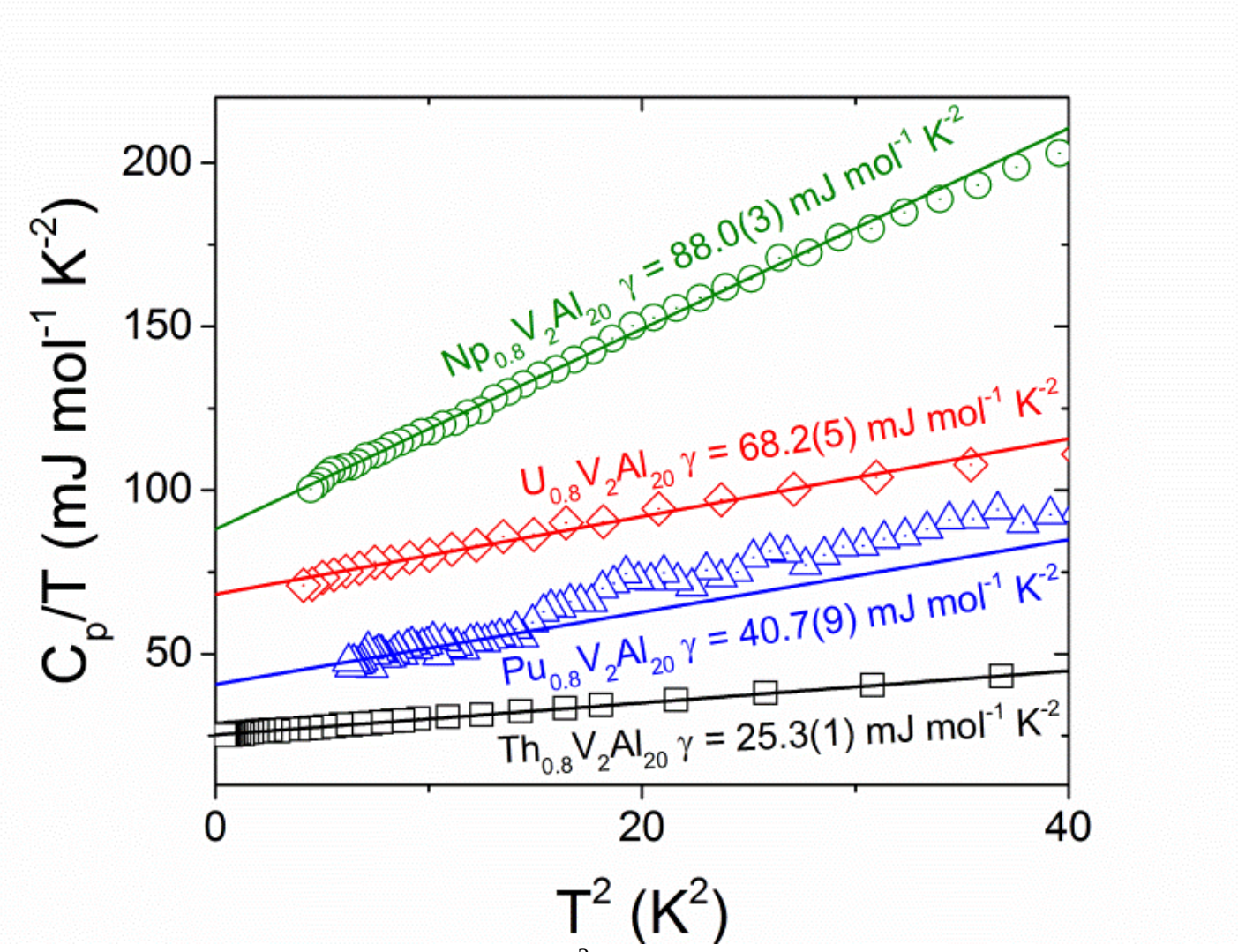

Fig. 4. Low-temperature specific heat of $A_{0.8}V_2Al_{20}$ presented as $C_p/T$ vs. $T^2$. Solid lines are linear fits to the experimental data. Specific heat of $Ce_{0.8}V_2Al_{20}$ was not shown for clarity but is presented in Fig. 5.

In a simple Debye model, the coefficient β is related to the Debye temperature $\Theta_D$ through the relation:

$$\Theta_D = \sqrt[3]{\frac{12\pi^4 nR}{5\beta}} \qquad (2)$$

where $n$ = 22.8 is the number of atoms per formula unit and $R$ is the gas constant. The so-derived Debye temperatures vary from 244 K for Np compound to 449 K for $Th_{0.8}V_2Al_{20}$ (see Table 2). The values of $\Theta_D$ are significantly smaller than those estimated for $RV_2Al_{20}$ ($R$ = Sc, Y, La, Lu), where they ranged from 502 K ($LuV_2Al_{20}$) to 536 K ($ScV_2Al_{20}$) [12], however in case of $Th_{0.8}V_2Al_{20}$ $\Theta_D$ is similar as in $ThCr_2Al_{20}$ (457 K) [34].

Table 2. Values of γ and β specific heat coefficients extracted from the linear fits to the $C_p/T$ vs. $T^2$ data (see Fig. 4). Numbers in parentheses indicate the statistical uncertainty of the least significant digit.

| $A$ in $A_{0.8}V_2Al_{20}$: | Ce | Th | U | Np | Pu |
|---|---|---|---|---|---|
| γ (mJ $mol^{-1}$ $K^{-2}$) | 28.7(1) | 25.3(1) | 68.2(5) | 88.0(3) | 40.7(9) |
| β (mJ $mol^{-1}$ $K^{-4}$) | 1.69(1) | 0.489(4) | 1.19(3) | 3.06(2) | 1.10(9) |
| $\Theta_D$ (K) | 297(1) | 449(1) | 334(3) | 244(1) | 343(9) |

The wide range of $\Theta_D$ obtained for isostructural compounds indicates that the phonon vibrations in $AV_2Al_{20}$ do not exhibit a quadratic frequency dependence assumed in the Debye model. We have recently reported that the low-energy optical modes affect significantly the low temperature lattice specific heat of $RV_2Al_{20}$ ($R$ = Sc, Lu) [12], resulting in a deviation from the linearity of the $C_p/T$ vs. $T^2$ relation well below the $\Theta_D/50$ limit. A similar situation likely arises in $Ce_{0.8}V_2Al_{20}$, $U_{0.8}V_2Al_{20}$ and $Np_{0.8}V_2Al_{20}$ for which a pronounced nonlinearity in the $C_p/T$ vs. $T^2$ curve is observed (see fig. 5). For these three compounds, the low-temperature specific heat data can be approximated by the formula

$$\frac{C_p}{T}(T^2) = \gamma_1 + \beta_1 T^2 + A \cdot \frac{C_E(T)}{T} \quad (3)$$

In the above equation, $C_E$ is the contribution of a low-energy optical mode described by the Einstein phonon specific heat model:

$$C_E(T) = 3nR\left(\frac{\Theta_E}{T}\right)^2 \exp\left(\frac{\Theta_E}{T}\right)\left(\exp\left(\frac{\Theta_E}{T} - 1\right)\right)^{-2} \quad (4)$$

where $\Theta_E$ is the Einstein temperature. The parameters derived for $A_{0.8}V_2Al_{20}$ ($A$ = Ce, U, Np) from fitting Eq. 1 and Eq. 3 to the experimental data in the relevant temperature ranges are given in Table 3.

Table 3. Values of the parameters derived from fitting the low-temperature specific heat data of $A_{0.8}V_2Al_{20}$. Fields written in bold correspond to the fitting with eq. 3.

| | $Ce_{0.8}V_2Al_{20}$ | $U_{0.8}V_2Al_{20}$ | $Np_{0.8}V_2Al_{20}$ |
|---|---|---|---|
| $\gamma$ (mJ mol$^{-1}$ K$^{-2}$) | 28.7(1) | 68.2(5) | 88.0(3) |
| **$\gamma_1$ (mJ mol$^{-1}$ K$^{-2}$)** | **31.1(2)** | **68.1(5)** | **93.9(4)** |
| $\beta$ (mJ mol$^{-1}$ K$^{-4}$) | 1.69(1) | 1.19(3) | 3.06(2) |
| $\Theta_D$ (K) | 297(1) | 334(3) | 244(1) |
| **$\beta_1$ (mJ mol$^{-1}$ K$^{-4}$)** | **1.01(1)** | **0.659(18)** | **1.78(2)** |
| **$\Theta_{D1}$ (K)** | **353(2)** | **407(4)** | **292(1)** |
| **B · $10^5$ (-)** | **0.78(3)** | **0.75(4)** | **2.21(7)** |
| **$\Theta_E$ (K)** | **17.0(2)** | **15.2(5)** | **20.1(2)** |

The Debye temperature found from fitting with Eq. 3 is 50-70 K higher than that obtained from a simple Debye fit (cf. Fig. 4). The new value of $\Theta_D$ for U-bearing compound is close to the one observed in $UNb_2Al_{20}$ (381 K) [36] and lower than in $UCr_2Al_{20}$ (474 K) [34]. It is also comparable with $\Theta_D$ estimated for $ErV_2Al_{20}$ (370 K) [50]. The observed Debye temperatures are most likely affected by the presence of low-energy Einstein modes. Such effect was observed in $Ga_xV_2Al_{20}$ where the appearance of the Einstein mode was correlated with a large drop in $\Theta_D$ [51].

It is also worthwhile noting that the obtained values of $\Theta_E$ are similar to $\Theta_E$ = 21 K estimated for the compound $VAl_{10.1}$ described as an 'Einstein solid' [52]. The $A$ prefactors are small, yet the curvatures of specific heat caused by the presence of low-energy modes are easily seen (see Fig. 5). Such Einstein mode but with twice higher $\Theta_E$ was also recently reported in $ErV_2Al_{20}$ single crystals [50].

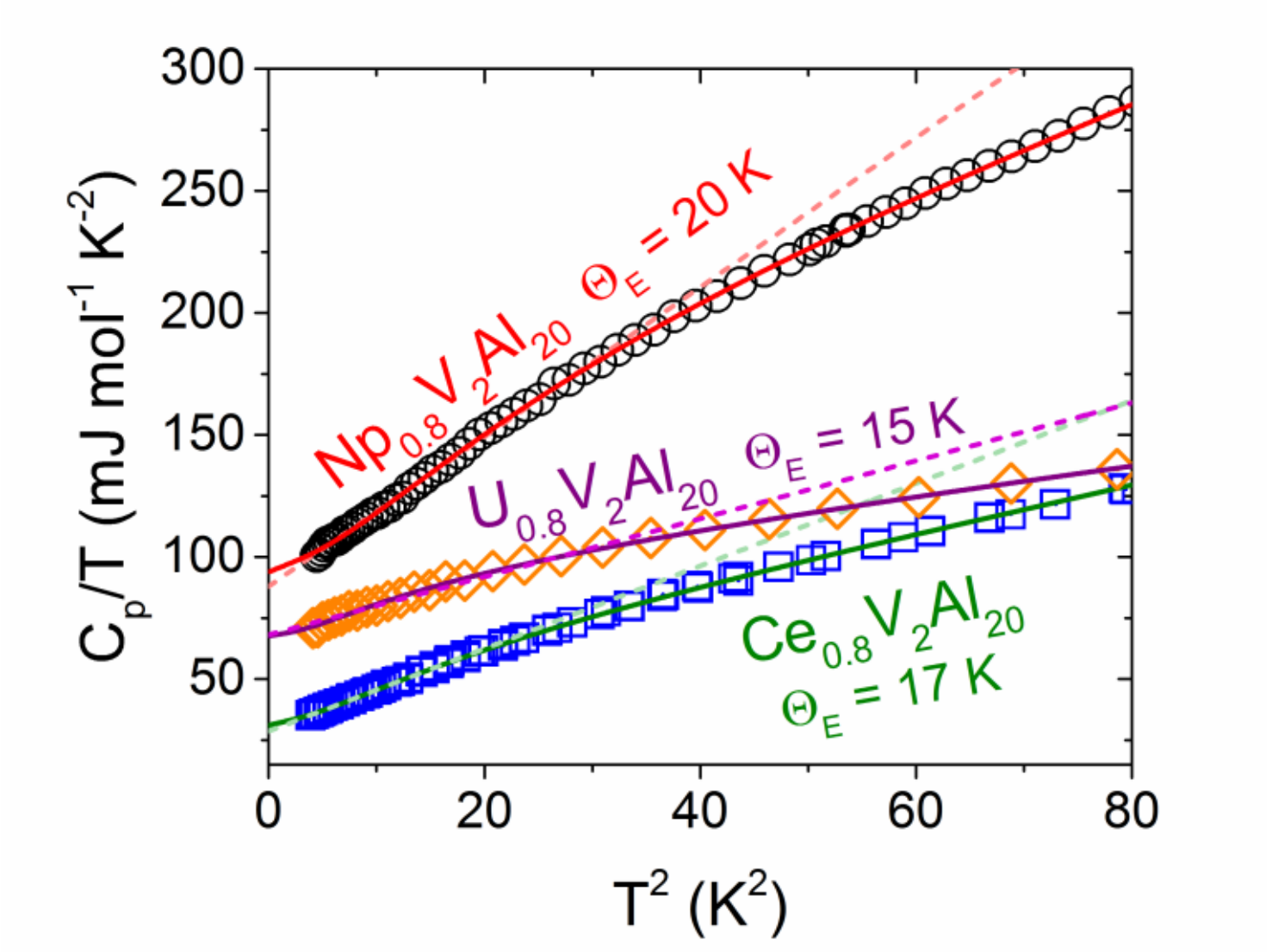

Fig. 5. Fits (solid lines) to the low temperature specific heat of $Ce_{0.8}V_2Al_{20}$ (blue squares), $U_{0.8}V_2Al_{20}$ (orange diamonds) and $Np_{0.8}V_2Al_{20}$ (black circles). Dashed lines show the linear fits (see Fig. 4).

The electronic heat capacity parameter $\gamma$ derived from fit was used to extract the lattice part of specific heat of Ce-, Th-, U- and Np-bearing compounds:

$$C_{lattice}(T) = C_p(T) - \gamma T \quad (5)$$

Figure 6 shows the lattice specific heat in the form of $C_{lattice}/T^3$ vs. log T plot in which the Einstein mode appears as a broad peak, position of which corresponds to $\Theta_E$ divided by a factor of ca. 5. The Einstein temperatures estimated in this way are in good agreement with those derived from specific heat fits, confirming correctness of our approach. A second Einstein mode with $\Theta_E \approx 130$ K, common for the four compounds is also seen in the plot. In a recent study the Einstein mode around 130-150 K was found in six $RT_2Al_{20}$ ($R$ = La, Eu, Gd, T = Ti, V) compounds [53], thus it is likely a common feature of all $CeCr_2Al_{20}$-type aluminides, however no lower Einstein modes were seen, at least for the La-bearing intermetallics. If the low-energy mode observed in Ce-, U- and Np-bearing samples is associated with the 8*a* atom, one would expect it to arise more likely in systems with small atoms in 8*a* position. This is in agreement with our observations, since Ce, U, and Np are smaller than La (see Fig. 2(a)). The low values of *B* prefactor, however, might suggest that the Einstein modes are associated with 8*a* site vacancies.

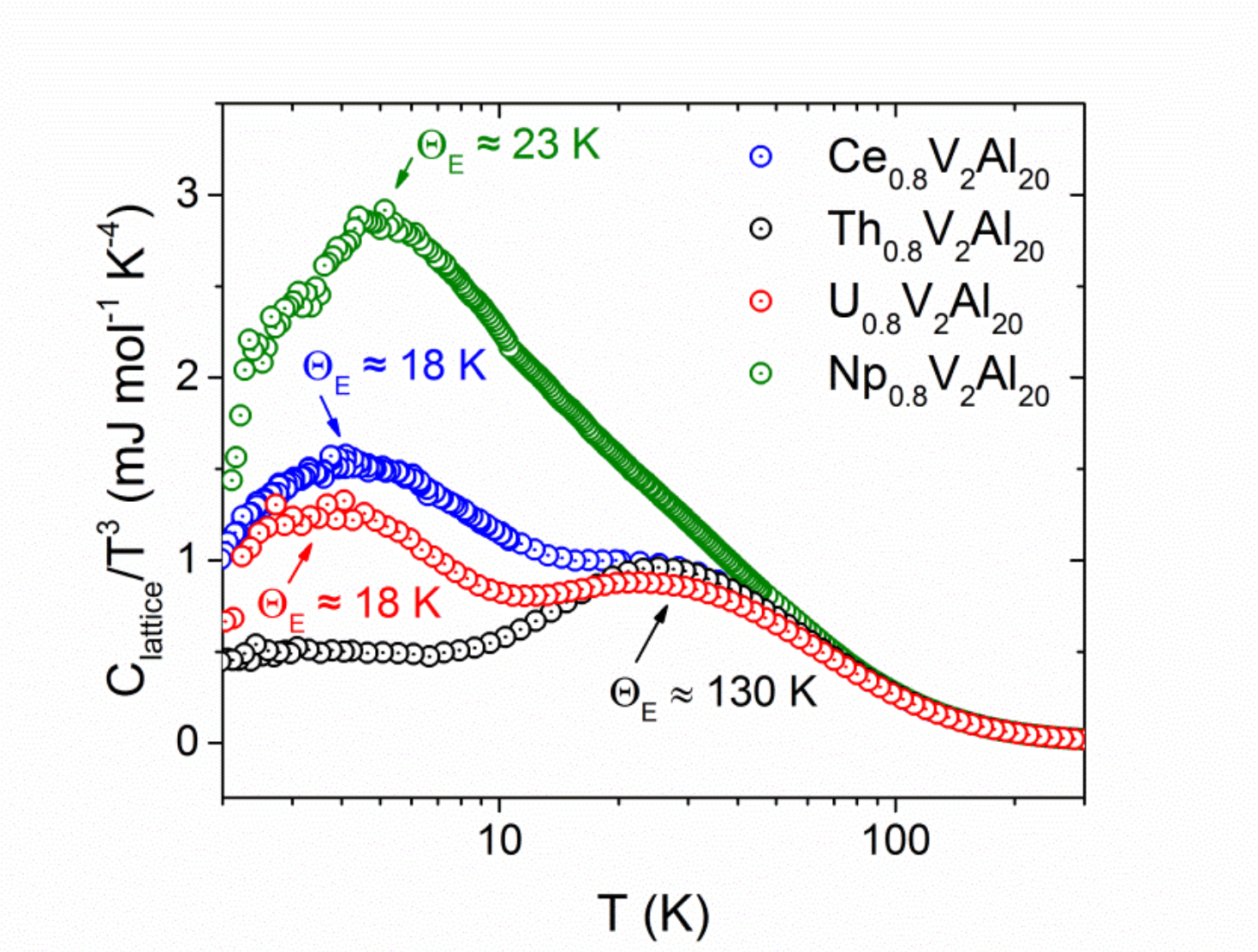

Fig. 6. Lattice specific heat ($C_{lattice}$) of $A_{0.8}V_2Al_{20}$ ($A$ = Ce, Th, U, Np). The broad peaks seen for Ce-, U- and Np-compounds at 4-5 K correspond to the low energy Einstein modes with $\Theta_E$ ca. 5 times larger than the peak position. Note also the second Einstein mode with $\Theta_E \approx 130$ K.

## 4. Conclusions

Four actinide intermetallic compounds, crystallizing in $CeCr_2Al_{20}$-type cage structure, were synthesized and characterized. Two of them, $Np_{0.8}V_2Al_{20}$ and $Pu_{0.8}V_2Al_{20}$ are reported for the first time. A weak yet negative correlation between the U atom content (x) in $U_xV_2Al_{20}$ and the lattice parameter was observed, contrary to the case of $Ce_xV_2Al_{20}$ and $Th_xV_2Al_{20}$ series. Moreover, the actinide-containing $A_{0.8}V_2Al_{20}$ compounds reveal a linear decrease in the lattice parameter with increasing the covalent radius of the $A$ element that is in sharp contrast with a positive, approximately linear relation seen in the lanthanide-bearing $MV_2Al_{20}$ materials. This behavior is likely caused by different character of 5*f* (mostly itinerant) and 4*f* (localized) electrons.

Magnetic susceptibility measurements showed that the actinide atoms in $A_{0.8}V_2Al_{20}$ ($A$ = U, Np, Pu) carry no local magnetic moment. The lack of magnetic moment was postulated to arise from a strong hybridization of the 5*f* electrons with the valence states of neighboring Al atoms. The values of the Sommerfeld coefficient extracted from the specific heat measurement varied significantly amidst the four compounds, ranging from 25 to 94 mJ mol$^{-1}$ K$^{-2}$ in Th- and Np-bearing sample, respectively. The low-temperature specific heat of $Ce_{0.8}V_2Al_{20}$, $U_{0.8}V_2Al_{20}$, and $Np_{0.8}V_2Al_{20}$ shows a contribution of a low-energy Einstein phonon mode ($\Theta_E \approx 17$, 15, and 20 K, respectively).

There are only two Np-containing superconductors known up to date, of which in only one ($NpPd_5Al_2$ [54,55]) the 5*f* states were found to contribute to the superconductivity [56]. Since it was shown previously that the $CeCr_2Al_{20}$ structure is favorable for superconductivity and that its presence is associated with the low-energy phonon modes of cage-filling atoms [8,9,12], $Np_{0.8}V_2Al_{20}$ is an excellent candidate for being a superconductor. Therefore measurements below 2 K are of high interest.

**Acknowledgements**

Authors gratefully acknowledge dr Rachel Eloirdi, Daniel Bouëxière, Pedro Amador-Celdran, dr Amir Hen, and Krisztina Varga from the Directorate for Nuclear Safety and Security, Karlsruhe for help with preparation and characterization of radioactive samples.

The research performed at the Gdansk University of Technology was financially supported by the National Science Centre (Poland) grant (DEC-2012/07/E/ST3/00584).

The access to the infrastructures of JRC-ITU and financial support provided by the European Commission within its Actinide User Laboratory program is acknowledged.

The high purity Np metal required for the fabrication of the compound was made available through a loan agreement between Lawrence Livermore National Laboratory and ITU, in the framework of a collaboration involving LLNL, Los Alamos National Laboratory, and the US Department of Energy.

**References**

[1] Y. Verbovytsky, K. Latka, K. Tomala, The crystal structure and magnetic properties of the $GdV_2Al_{20}$ and $GdCr_2Al_{20}$ ternary compounds, J. Alloys Compd. 442 (2007) 334–336. doi:10.1016/j.jallcom.2006.07.148.

[2] A. Sakai, S. Nakatsuji, Strong valence fluctuation effects in $SmTr_2Al_{20}$ (Tr=Ti, V, Cr), Phys. Rev. B. 84 (2011) 201106. doi:10.1103/PhysRevB.84.201106.

[3] N. Kase, Y. Shimura, S. Kittaka, T. Sakakibara, S. Nakatsuji, T. Nakano, N. Takeda, J. Akimitsu, Antiferromagnetic transition of the caged compound $TmTi_2Al_{20}$, J. Phys. Conf. Ser. 592 (2015) 12052. doi:10.1088/1742-6596/592/1/012052.

[4] P. Swatek, D. Kaczorowski, Magnetic properties of $EuCr_2Al_{20}$, J. Magn. Magn. Mater. 416 (2016) 348–352. doi:10.1016/j.jmmm.2016.04.086.

[5] C.H. Wang, J.M. Lawrence, E.D. Bauer, K. Kothapalli, J.S. Gardner, F. Ronning, K. Gofryk, J.D. Thompson, H. Nakotte, F. Trouw, Unusual signatures of the ferromagnetic transition in the heavy fermion compound $UMn_2Al_{20}$, Phys. Rev. B. 82 (2010) 94406. doi:10.1103/PhysRevB.82.094406.

[6] P. Wiśniewski, P. Swatek, A. Gukasov, D. Kaczorowski, Ferromagnetism in $UMn_2Al_{20}$ studied with polarized neutron diffraction and bulk magnetic measurements, Phys. Rev. B. 86 (2012) 54438. doi:10.1103/PhysRevB.86.054438.

[7] T. Namiki, Q. Lei, Y. Isikawa, K. Nishimura, Possible Heavy Fermion State of the Caged Cubic Compound NdV2Al20, J. Phys. Soc. Jpn. 85 (2016) 73706. doi:10.7566/JPSJ.85.073706.

[8] A. Onosaka, Y. Okamoto, J. Yamaura, Z. Hiroi, Superconductivity in the Einstein Solid $A_xV_2Al_{20}$ (A= Al and Ga), J. Phys. Soc. Jpn. 81 (2012) 23703. doi:10.1143/JPSJ.81.023703.

[9] T. Klimczuk, M. Szlawska, D. Kaczorowski, J.R. O'Brien, D.J. Safarik, Superconductivity in the Einstein solid V $Al_{10.1}$, J. Phys. Condens. Matter. 24 (2012) 365701. doi:10.1088/0953-8984/24/36/365701.

[10] A. Sakai, K. Kuga, S. Nakatsuji, Superconductivity in the Ferroquadrupolar State in the Quadrupolar Kondo Lattice $PrTi_2Al_{20}$, J. Phys. Soc. Jpn. 81 (2012) 83702. doi:10.1143/JPSJ.81.083702.

[11] M. Tsujimoto, Y. Matsumoto, T. Tomita, A. Sakai, S. Nakatsuji, Heavy-Fermion Superconductivity in the Quadrupole Ordered State of $PrV_2Al_{20}$, Phys. Rev. Lett. 113 (2014) 267001. doi:10.1103/PhysRevLett.113.267001.

[12] M.J. Winiarski, B. Wiendlocha, M. Sternik, P. Wiśniewski, J.R. O'Brien, D. Kaczorowski, T. Klimczuk, Rattling-enhanced superconductivity in $MV_2Al_{20}$ (M = Sc, Lu, Y) intermetallic cage compounds, Phys. Rev. B. 93 (2016) 134507. doi:10.1103/PhysRevB.93.134507.

[13] S. Niemann, W. Jeitschko, Ternary Aluminides $AT_2Al_{20}$ (A = Rare Earth Elements and Uranium; T = Ti, Nb, Ta, Mo, and W) with CeCr2Al20-Type Structure, J. Solid State Chem. 114 (1995) 337–341. doi:10.1006/jssc.1995.1052.

[14] V.M.T. Thiede, W. Jeitschko, S. Niemann, T. Ebel, $EuTa_2Al_{20}$, $Ca_6W_4Al_{43}$ and other compounds with $CeCr_2Al_{20}$ and $Ho_6Mo_4Al_{43}$ type structures and some magnetic properties of these compounds, J. Alloys Compd. 267 (1998) 23–31. doi:10.1016/S0925-8388(97)00532-X.

[15] K. Momma, F. Izumi, *VESTA 3* for three-dimensional visualization of crystal, volumetric and morphology data, J. Appl. Crystallogr. 44 (2011) 1272–1276. doi:10.1107/S0021889811038970.

[16] I. Halevy, E. Sterer, M. Aizenshtein, G. Kimmel, D. Regev, E. Yahel, L.C.J. Pereira, A.P. Goncalves, High pressure studies of a new ternary actinide compound, $UV_2Al_{20}$, J. Alloys Compd. 319 (2001) 19–21. doi:10.1016/S0925-8388(01)00881-7.
[17] A. Uziel, A.I. Bram, A. Venkert, A.E. Kiv, D. Fuks, L. Meshi, Abrupt symmetry decrease in the $ThT_2Al_{20}$ alloys (T = 3d transition metal), J. Alloys Compd. 648 (2015) 353–359. doi:10.1016/j.jallcom.2015.06.216.
[18] H. Okamoto, Al-Np (Aluminum-Neptunium), J. Phase Equilibria Diffus. 33 (2012) 488–488. doi:10.1007/s11669-012-0088-y.
[19] H. Okamoto, Al-Pu (Aluminum-Plutonium), J. Phase Equilibria Diffus. 30 (2009) 293–294. doi:10.1007/s11669-009-9506-1.
[20] H.M. Rietveld, A profile refinement method for nuclear and magnetic structures, J. Appl. Crystallogr. 2 (1969) 65–71. doi:10.1107/S0021889869006558.
[21] J. Rodríguez-Carvajal, Recent advances in magnetic structure determination by neutron powder diffraction, Phys. B Condens. Matter. 192 (1993) 55–69. doi:10.1016/0921-4526(93)90108-I.
[22] T. Klimczuk, A.B. Shick, R. Springell, H.C. Walker, A.H. Hill, E. Colineau, J.-C. Griveau, D. Bouëxière, R. Eloirdi, R. Caciuffo, Bulk properties and electronic structure of PuFeAsO, Phys. Rev. B. 86 (2012) 174510. doi:10.1103/PhysRevB.86.174510.
[23] A.C. Walters, H.C. Walker, R. Springell, M. Krisch, A. Bosak, A.H. Hill, C.E. Zvorişte-Walters, E. Colineau, J-C Griveau, D. Bouëxière, R. Eloirdi, R. Caciuffo, T. Klimczuk, Absence of superconductivity in fluorine-doped neptunium pnictide NpFeAsO, J. Phys. Condens. Matter. 27 (2015) 325702. doi:10.1088/0953-8984/27/32/325702.
[24] P. Javorský, F. Wastin, E. Colineau, J. Rebizant, P. Boulet, G. Stewart, Low-temperature heat capacity measurements on encapsulated transuranium samples, J. Nucl. Mater. 344 (2005) 50–55. doi:10.1016/j.jnucmat.2005.04.015.
[25] P.J. Brown, The structure of the intermetallic phase α' (VAl), Acta Crystallogr. 12 (1959) 995–1002. doi:10.1107/S0365110X59002821.
[26] J.L. Smith, E.A. Kmetko, Magnetism or bonding: A nearly periodic table of transition elements, J. Common Met. 90 (1983) 83–88. doi:10.1016/0022-5088(83)90119-4.
[27] O. Moze, L.D. Tung, J.J.M. Franse, K.H.J. Buschow, Crystal structure and magnetic properties of $CeV_2Al_{20}$ and $CeCr_2Al_{20}$, J. Alloys Compd. 268 (1998) 39–41. doi:10.1016/S0925-8388(97)00586-0.
[28] M.J. Kangas, D.C. Schmitt, A. Sakai, S. Nakatsuji, J.Y. Chan, Structure and physical properties of single crystal $PrCr_2Al_{20}$ and $CeM_2Al_{20}$ (M=V, Cr): A comparison of compounds adopting the $CeCr_2Al_{20}$ structure type, J. Solid State Chem. 196 (2012) 274–281. doi:10.1016/j.jssc.2012.06.035.
[29] W.H. Zachariasen, Metallic radii and electron configurations of the 5f–6d metals, J. Inorg. Nucl. Chem. 35 (1973) 3487–3497. doi:10.1016/0022-1902(73)80357-4.
[30] A.C. Lawson, 5f-electron localization in the actinide metals: thorides, actinides and the Mott transition, Philos. Mag. Lett. 96 (2016) 85–89. doi:10.1080/09500839.2016.1157634.
[31] B. Cordero, V. Gómez, A.E. Platero-Prats, M. Revés, J. Echeverría, E. Cremades, F. Barragán, S. Alvarez, Covalent radii revisited, Dalton Trans. (2008) 2832–2838. doi:10.1039/B801115J.
[32] B.H. Toby, R factors in Rietveld analysis: How good is good enough?, Powder Diffr. 21 (2006) 67–70. doi:10.1154/1.2179804.
[33] J. Huot, R. Černý, Neutron Powder Diffraction, in: H. Fritzsche, J. Huot, D. Fruchart (Eds.), Neutron Scatt. Nucl. Tech. Hydrog. Mater., Springer International Publishing, 2016: pp. 31–89. http://link.springer.com/chapter/10.1007/978-3-319-22792-4_3.
[34] P. Swatek, D. Kaczorowski, Magnetic and electrical properties of $UCr_2Al_{20}$ single crystals, J. Solid State Chem. 191 (2012) 191–194. doi:10.1016/j.jssc.2012.03.018.
[35] Y. Matsumoto, T.D. Matsuda, N. Tateiwa, E. Yamamoto, Y. Haga, Z. Fisk, Single crystal growth and physical properties of $UT_2Al_{20}$ (T=Transition Metal), J. Korean Phys. Soc. 63 (2013) 363–366. doi:10.3938/jkps.63.363.
[36] C. Moussa, A. Berche, M. Pasturel, J. Barbosa, B. Stepnik, S. Dubois, O. Tougait, The U-Nb-Al ternary system: experimental and simulated investigations of the phase equilibria and study of the crystal structure and electronic properties of the intermediate phases, J. Alloys Compd. 691 (2017) 893–905. doi:10.1016/j.jallcom.2016.08.257.
[37] D.D. Koelling, B.D. Dunlap, G.W. Crabtree, f-electron hybridization and heavy-fermion compounds, Phys. Rev. B. 31 (1985) 4966–4971. doi:10.1103/PhysRevB.31.4966.
[38] P. Swatek, D. Kaczorowski, Intermediate valence behavior in the novel cage compound $CeIr_2Zn_{20}$, J. Phys. Condens. Matter. 25 (2013) 55602. doi:10.1088/0953-8984/25/5/055602.
[39] B.D. White, D. Yazici, P.-C. Ho, N. Kanchanavatee, N. Pouse, Y. Fang, A.J. Breindel, A.J. Friedman, M.B. Maple, Weak hybridization and isolated localized magnetic moments in the compounds $CeT_2Cd_{20}$ (T = Ni, Pd), J. Phys. Condens. Matter. 27 (2015) 315602. doi:10.1088/0953-8984/27/31/315602.
[40] A. Labarta, H. Doto, Y. Hirose, F. Honda, D. Li, Y. Homma, D. Aoki, R. Settai, Single Crystal Growth and Electronic State of $UPd_2Cd_{20}$, Phys. Procedia. 75 (2015) 56–61. doi:10.1016/j.phpro.2015.12.009.
[41] P. Swatek, D. Kaczorowski, Heavy Fermion Behavior in $UT_2Zn_{20}$ (T = Fe, Co, Ru, Rh, Ir) Compounds, J. Phys. Soc. Jpn. 80SA (2011) SA106. doi:10.1143/JPSJS.80SA.SA106.

[42] J. Gal, I. Yaar, S. Fredo, I. Halevy, W. Potzel, S. Zwirner, G.M. Kalvius, Magnetic and electronic properties of cubic $NpX_3$ intermetallics, Phys. Rev. B. 46 (1992) 5351–5356. doi:10.1103/PhysRevB.46.5351.
[43] H. Yamagami, Electronic Structure and Fermi Surface of $NpGe_3$, J. Phys. Soc. Jpn. 75 (2006) 24711. doi:10.1143/JPSJ.75.024711.
[44] W.J. Nellis, A.R. Harvey, M.B. Brodsky, Stabilization of the 5f Energy Band in Actinide- $Rh_3$ Intermetallic Compounds, in: AIP Conf. Proc., AIP Publishing, 1973: pp. 1076–1080. doi:10.1063/1.2946747.
[45] D. Aoki, H. Yamagami, Y. Homma, Y. Shiokawa, E. Yamamoto, A. Nakamura, Y. Haga, R. Settai, Y. Ōnuki, Itinerant 5 f Electrons and the Fermi Surface Properties in an Enhanced Pauli Paramagnet $NpGe_3$, J. Phys. Soc. Jpn. 74 (2005) 2149–2152. doi:10.1143/JPSJ.74.2149.
[46] E. Colineau, P. Javorský, P. Boulet, F. Wastin, J.C. Griveau, J. Rebizant, J.P. Sanchez, G.R. Stewart, Magnetic and electronic properties of the antiferromagnet $NpCoGa_5$, Phys. Rev. B. 69 (2004) 184411. doi:10.1103/PhysRevB.69.184411.
[47] T. Klimczuk, J.-C. Griveau, P.Gaczynski, R. Eloirdi, E. Colineau, R. Caciuffo, Crystal structure and physical properties of $NpRh_2Sn$, a new Np-based ternary compound, J. Phys. Conf. Ser. 273 (2011) 12024. doi:10.1088/1742-6596/273/1/012024.
[48] T. Klimczuk, A.B. Shick, A.L. Kozub, J.-C. Griveau, E. Colineau, M. Falmbigl, F. Wastin, P. Rogl, Ferro- and antiferromagnetism in (Np, Pu)BC, APL Mater. 3 (2015) 41803. doi:10.1063/1.4913564.
[49] A. Hen, N. Magnani, J.-C. Griveau, R. Eloirdi, E. Colineau, J.-P. Sanchez, I. Halevy, A.L. Kozub, A.B. Shick, I. Orion, R. Caciuffo, Site-selective magnetic order of neptunium in $Np_2Ni_{17}$, Phys. Rev. B. 92 (2015) 24410. doi:10.1103/PhysRevB.92.024410.
[50] M.J. Winiarski, T. Klimczuk, Crystal structure and low-energy Einstein mode in $ErV_2Al_{20}$ intermetallic cage compound, J. Solid State Chem. 245 (2017) 10–16. doi:10.1016/j.jssc.2016.09.029.
[51] Z. Hiroi, A. Onosaka, Y. Okamoto, J. Yamaura, H. Harima, Rattling and Superconducting Properties of the Cage Compound $Ga_xV_2Al_{20}$, J. Phys. Soc. Jpn. 81 (2012) 124707. doi:10.1143/JPSJ.81.124707.
[52] D.J. Safarik, T. Klimczuk, A. Llobet, D.D. Byler, J.C. Lashley, J.R. O'Brien, N.R. Dilley, Localized anharmonic rattling of Al atoms in $VAl_{10.1}$, Phys. Rev. B. 85 (2012) 14103. doi:10.1103/PhysRevB.85.014103.
[53] K.R. Kumar, H.S. Nair, R. Christian, A. Thamizhavel, A.M. Strydom, Magnetic, specific heat and electrical transport properties of Frank–Kasper cage compounds $RTM_2Al_{20}$ [R = Eu, Gd and La; TM = V, Ti], J. Phys. Condens. Matter. 28 (2016) 436002. doi:10.1088/0953-8984/28/43/436002.
[54] D. Aoki, Y. Haga, T. D. Matsuda, N. Tateiwa, S. Ikeda, Y. Homma, H. Sakai, Y. Shiokawa, E. Yamamoto, A. Nakamura, R. Settai, Y. Ōnuki, Unconventional Heavy-Fermion Superconductivity of a New Transuranium Compound $NpPd_5Al_2$, J. Phys. Soc. Jpn. 76 (2007) 63701. doi:10.1143/JPSJ.76.063701.
[55] J.-C. Griveau, K. Gofryk, J. Rebizant, Transport and magnetic properties of the superconductor $NpPd_5Al_2$, Phys. Rev. B. 77 (2008) 212502. doi:10.1103/PhysRevB.77.212502.
[56] J.-C. Griveau, É. Colineau, Superconductivity in transuranium elements and compounds, Comptes Rendus Phys. 15 (2014) 599–615. doi:10.1016/j.crhy.2014.07.001.

# Synthesis and properties of $A_xV_2Al_{20}$ ($A$ = Ce, Th, U, Np, Pu) ternary actinide aluminides.

M. J. Winiarski[a],[*] J.-C. Griveau[b], E. Colineau[b], K. Wochowski[c], P. Wiśniewski[c],

D. Kaczorowski[c], R. Caciuffo[b], T. Klimczuk[a],[†],

[a] *Faculty of Applied Physics and Mathematics, Gdansk University of Technology, Narutowicza 11/12, 80-233 Gdansk, Poland*
[b] *European Commission, Joint Research Centre (JRC), Directorate for Nuclear Safety and Security, Postfach 2340, 76125 Karlsruhe, Germany*
[c] *Institute of Low Temperature and Structure Research, Polish Academy of Sciences, PO Box 1410, 50-950 Wrocław, Poland*

* *corresponding author: mwiniarski@mif.pg.gda.pl*

† *-corresponding author: tomasz.klimczuk@pg.gda.pl*

# Supplementary material

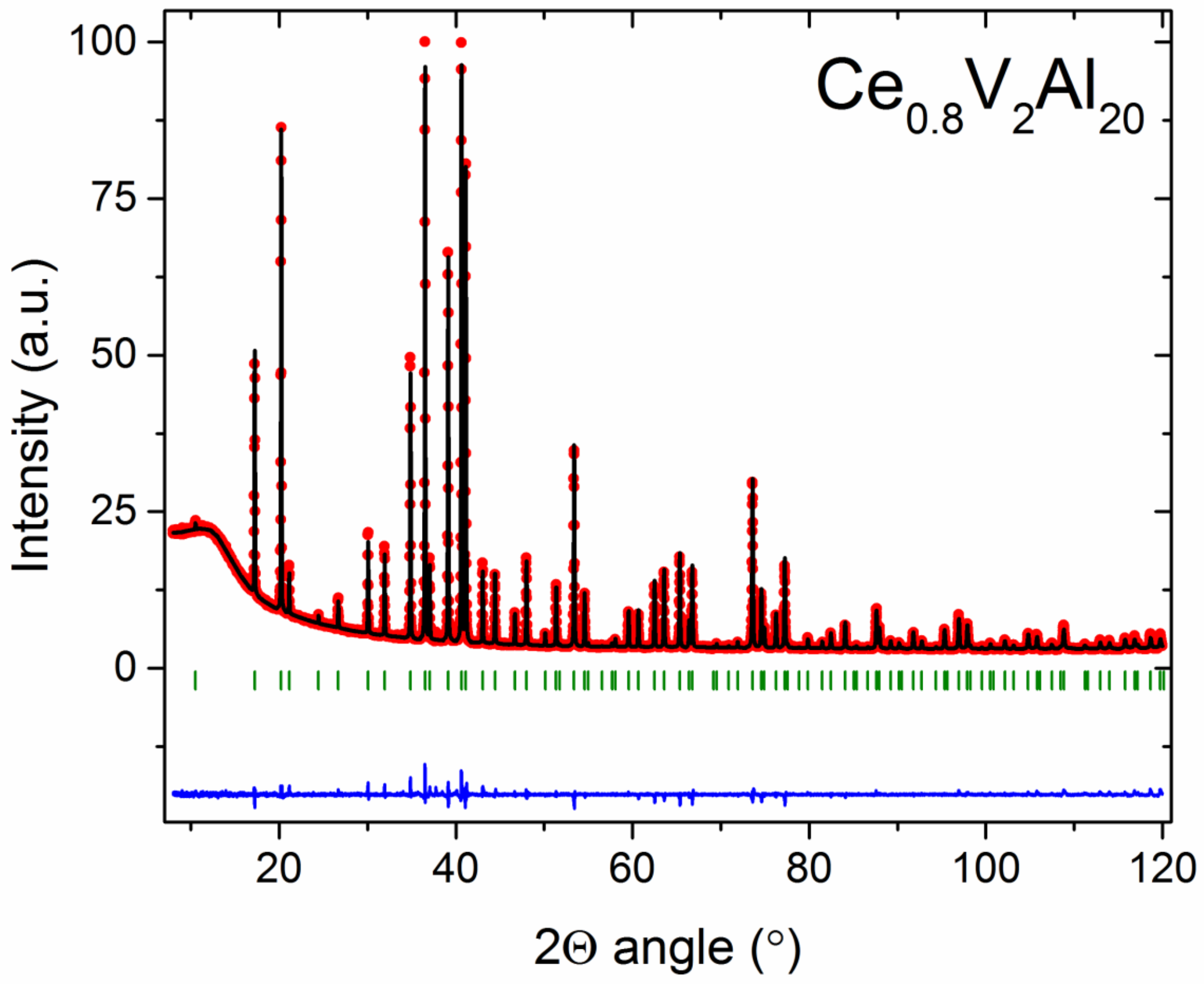

**Fig. S1 Rietveld fit (black solid line) to experimental powder x-ray diffraction pattern of $Ce_{0.8}V_2Al_{20}$ (red circles). Blue line shows a difference between the observed and calculated intensities. Green ticks mark the expected positions of Bragg reflections. Weak reflections from an unrecognized impurity phase(s) were observed at 2Θ ≈ 34.0, 37.7, 38.4 and 40.1°.**

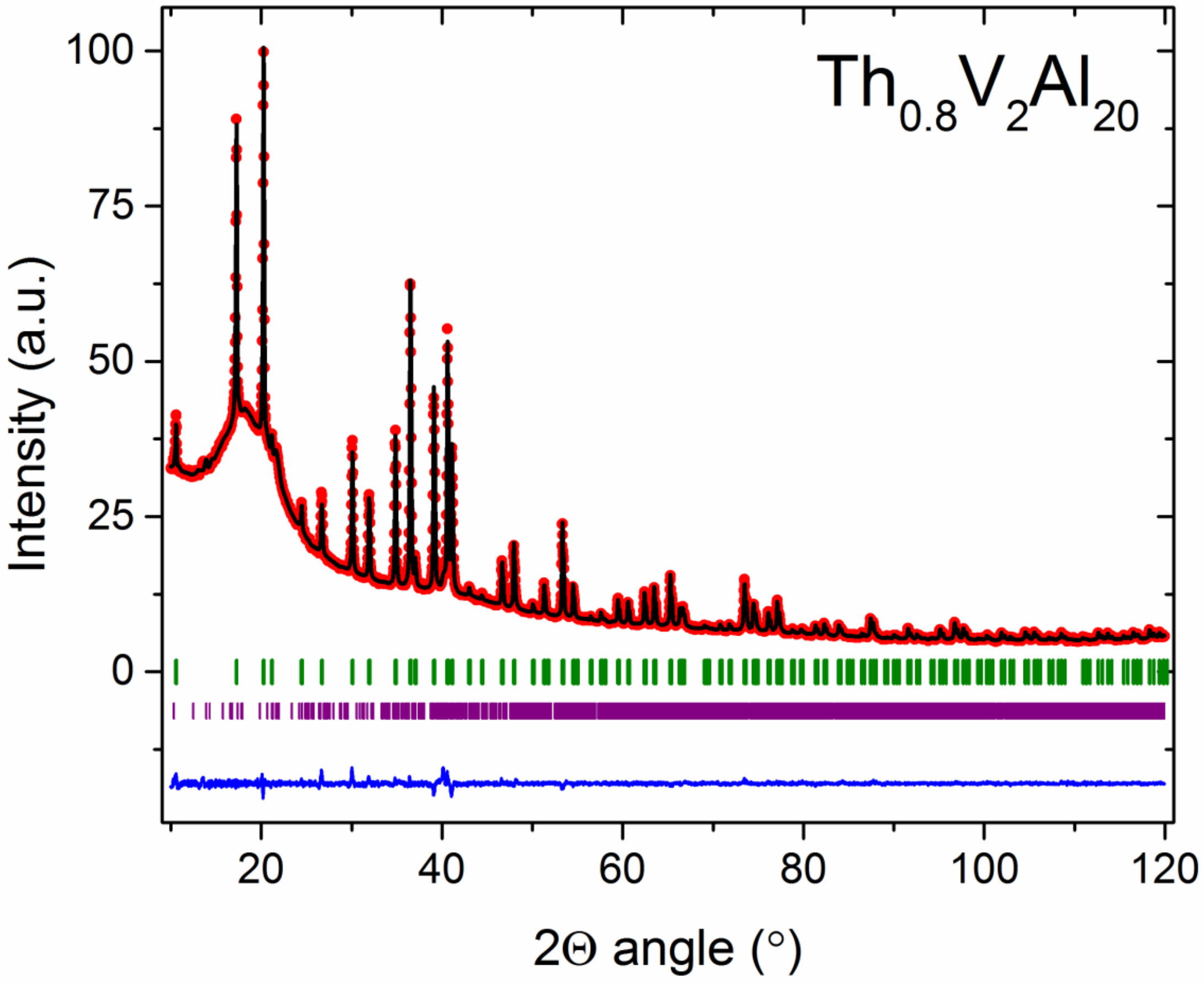

**Fig. S2 Rietveld fit (black solid line) to experimental powder x-ray diffraction pattern of $Th_{0.8}V_2Al_{20}$ (red circles). Blue line shows a difference between the observed and calculated intensities. Green ticks mark the expected positions of Bragg reflections. Impurity reflections were assigned to a $V_7Al_{45}$ binary phase and fit by profile matching. Purple ticks mark the reflections of the impurity phase. The resulting lattice parameters of $V_7Al_{45}$ (*a* = 25.62 Å, *b* = 7.63 Å, *c* = 11.09 Å, *β* = 129°) agree well with previously reported values [1].**

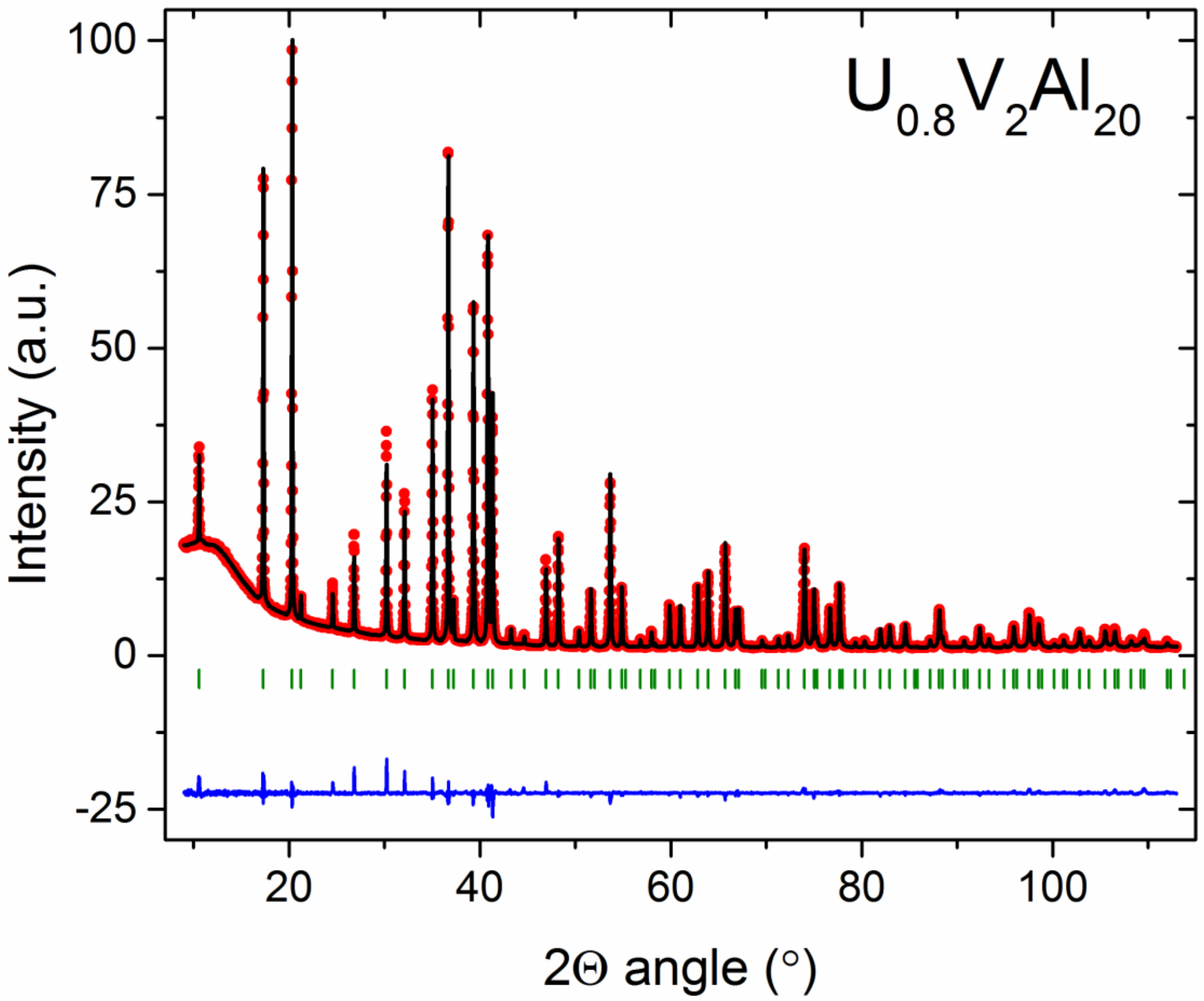

**Fig. S3 Rietveld fit (black solid line) to experimental powder x-ray diffraction pattern of $U_{0.8}V_2Al_{20}$ (red circles). Blue line shows a difference between the observed and calculated intensities. Green ticks mark the expected positions of Bragg reflections.**

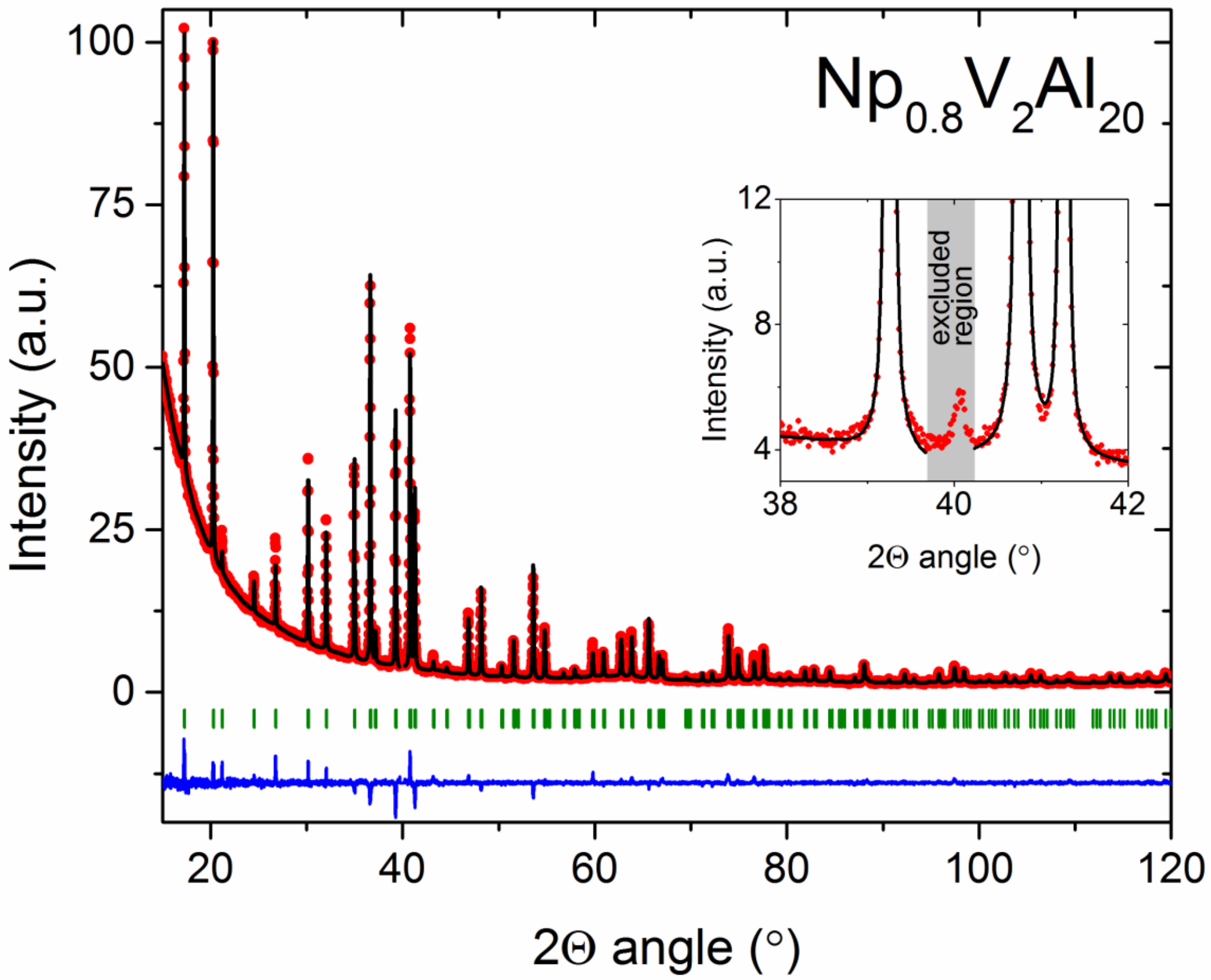

**Fig. S4 Rietveld fit (black solid line) to experimental powder x-ray diffraction pattern of $Np_{0.8}V_2Al_{20}$ (red circles). Blue line shows a difference between the observed and calculated intensities. Green ticks mark the expected positions of Bragg reflections. Inset shows the region excluded from the refinement due to presence of an unassigned impurity peak.**

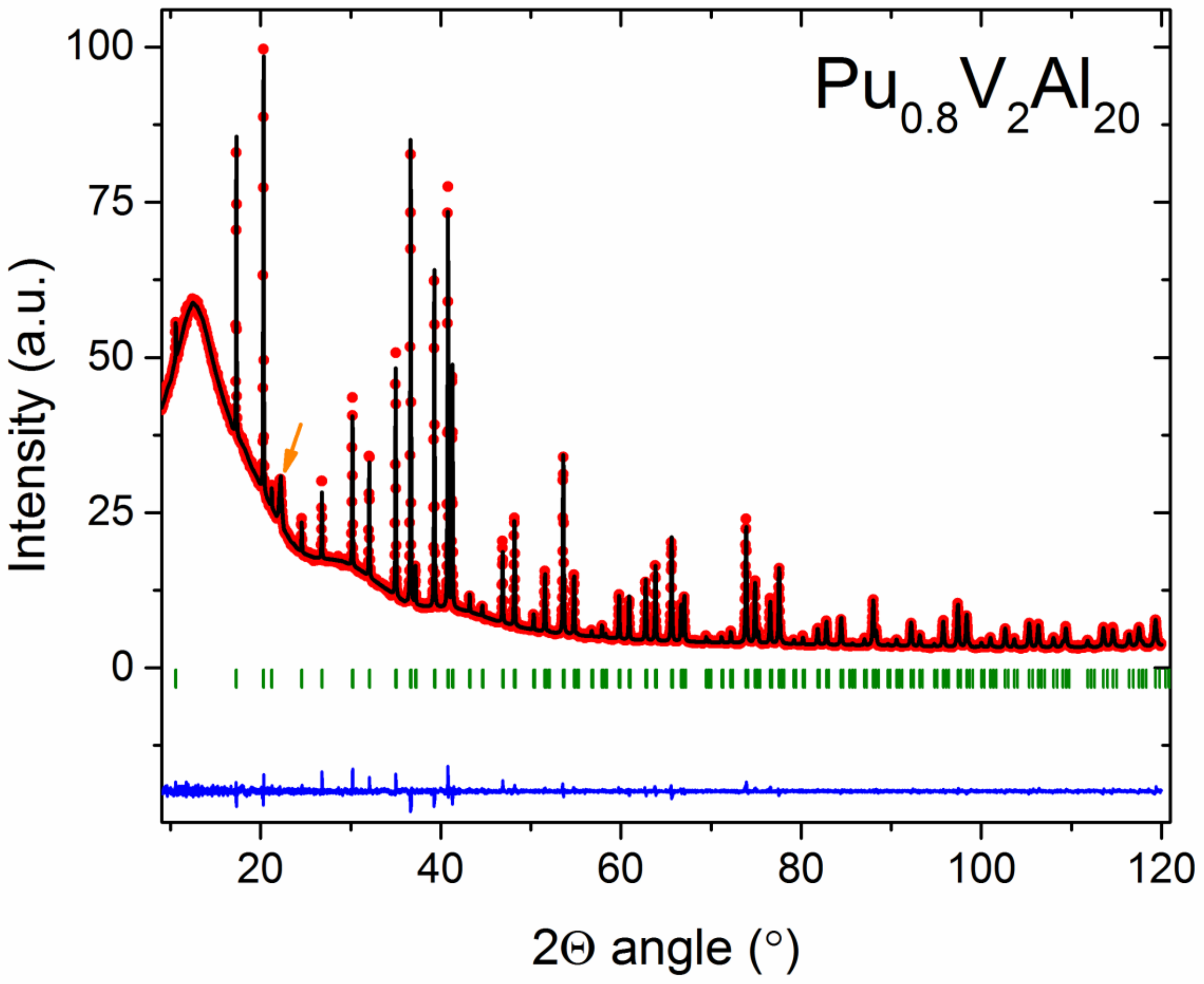

**Fig. S5 Rietveld fit (black solid line) to experimental powder x-ray diffraction pattern of $Pu_{0.8}V_2Al_{20}$ (red circles). Blue line shows a difference between the observed and calculated intensities. Green ticks mark the expected positions of Bragg reflections. Orange arrow marks a broad peak originating from a resin used to encapsulate the sample.**

**Table S1 Rietveld fit reliability factors (without background correction) for points with Bragg contributions for $A_{0.8}V_2Al_{20}$. For reliability factors corrected for background see Table 1.**

| $A$ = | Ce | Th | U | Np | Pu |
|---|---|---|---|---|---|
| $R_p$ (%) | 2.54 | 1.16 | 3.85 | 3.46 | 1.60 |
| $R_{wp}$ (%) | 3.47 | 1.54 | 5.20 | 4.94 | 2.22 |
| $R_{exp}$ (%) | 1.88 | 1.18 | 2.49 | 3.36 | 1.69 |
| $\chi^2$ | 3.40 | 1.71 | 4.34 | 2.16 | 1.73 |

## References

[1] P.J. Brown, The structure of the intermetallic phase α' (VAl), Acta Crystallogr. 12 (1959) 995–1002. doi:10.1107/S0365110X59002821.